\documentclass[sigconf,authorversion,nonacm]{acmart}

\usepackage{cite}
\usepackage{balance}
\usepackage{soul}
\usepackage[official]{eurosym}

\AtBeginDocument{%
  \providecommand\BibTeX{{%
    \normalfont B\kern-0.5em{\scshape i\kern-0.25em b}\kern-0.8em\TeX}}}

\setcopyright{acmcopyright}
\copyrightyear{2018}
\acmYear{2018}
\acmDOI{10.1145/1122445.1122456}

\newcommand{\parag}[1]{\noindent\textbf{#1. }}

\acmConference[Woodstock '18]{Woodstock '18: ACM Symposium on Neural
  Gaze Detection}{June 03--05, 2018}{Woodstock, NY}
\acmBooktitle{Woodstock '18: ACM Symposium on Neural Gaze Detection,
  June 03--05, 2018, Woodstock, NY}
\acmPrice{15.00}
\acmISBN{978-1-4503-XXXX-X/18/06}

\renewcommand{\cite}{\citep}

\begin{document}

\title{MiniV2G: An Electric Vehicle Charging Emulator}

\author{Luca Attanasio}
\orcid{0000-0003-3653-7080}
\email{luca.attanasio@studenti.unipd.it}
\affiliation{%
  \institution{Department of Mathematics\\University of Padua}
  \city{Padua}
  \country{Italy}
}

\author{Mauro Conti}
\orcid{0000-0002-3612-1934}
\email{conti@math.unipd.it}
\affiliation{%
  \institution{Department of Mathematics\\University of Padua}
  \city{Padua}
  \country{Italy}
}

\author{Denis Donadel}
\orcid{0000-0002-7050-9369}
\email{denis.donadel@studenti.unipd.it}
\affiliation{%
  \institution{Department of Mathematics\\University of Padua}
  \city{Padua}
  \country{Italy}
}

\author{Federico Turrin}
\orcid{0000-0001-5660-2447}
\email{turrin@math.unipd.it}
\affiliation{%
  \institution{Department of Mathematics\\University of Padua}
  \city{Padua}
  \country{Italy}
}


\begin{abstract}
The impact of global warming and the imperative to limit climate change have stimulated the need to develop new solutions based on renewable energy sources. One of the emerging trends in this endeavor are the Electric Vehicles (EVs), which use electricity instead of traditional fossil fuels as a power source, relying on the Vehicle-to-Grid (V2G) paradigm. 
The novelty of such a paradigm requires careful analysis to avoid malicious attempts. An attacker can exploit several surfaces, such as the remote connection between the Distribution Grid and Charging Supply or the authentication system between the charging Supply Equipment and the Electric Vehicles.
However, V2G architecture's high cost and complexity in implementation can restrain this field's research capability. 

In this paper, we approach this limitation by proposing MiniV2G, an open-source emulator to simulate Electric Vehicle Charging (EVC) built on top of Mininet and RiseV2G. MiniV2G is particularly suitable for security researchers to study and test real V2G charging scenarios. MiniV2G can reproduce with high fidelity a V2G architecture to easily simulate an EV charging process.
Finally, we present a MiniV2G application and show how MiniV2G can be used to study V2G communication and develop attacks and countermeasures that can be applied to real systems. Since we believe our tool can be of great help for research in this field, we also made it freely available.
\end{abstract}



\maketitle

\section{Introduction}
To face global warming, governments, companies, and researchers are increasingly endeavoring to reduce the daily human carbon footprint. The growing concern about this global catastrophe led to a worldwide movement with the aim to create a green and sustainable future.
The United States Environmental Protection Agency (EPA) in 2018 estimated that the larger percentage of Greenhouse Gas Emissions in the US (i.e., the $28.2\%$) is due to the transportation sector~\cite{web:car_pollution}.
To reduce the devastating impact of such a phenomenon, the transportation systems and automotive industries strive to replace traditional fossil fuels with renewable energy sources, such as electric power. The cars basing their power supplies on electricity are called Electric Vehicles (EVs). EVs market is steadily growing, and several companies predict that in 2040 EVs will account for more than 50\% of car Ecosystem~\cite{web:ev_market, web:ev_market_2}. However, charging all EVs requires a huge amount of energy. Companies are increasingly building infrastructures to satisfy this expanding power demand to manage the need for additional power supply. The paradigm defining EV power supply is called Vehicle-To-Grid (V2G). V2G defines the communication between the EV, asking for charging requests, and the Smart Grid, which delivers the energy. Furthermore, to require the recharging process, the user needs to authenticate the car within the Charging Station (CS) and pay for the required energy.
Nevertheless, this paradigm's novelty, along with its complexity and users' sensitive data involved, has led the way to potential risks and vulnerabilities exploitable by attackers. Moreover, the V2G communication relies on both network communication (e.g., user authentication) and physical communication (e.g., charging process), exposing the V2G paradigm to all Cyber-Physical Systems (CPS) known threats~\cite{humayed2017cyber}.
For instance, the authors of~\cite{baker2019losing} were able to sniff the Media Access Control (MAC) address from an unshielded charging cable, implying a potential breach of privacy.
Several other security weaknesses of the electric vehicle charging environment can be exploited to perform an array of different attacks: for example, a Denial of Service (DoS) can be employed to stop the charging station in a certain area, while an impersonation attack can be exploited to charge other users for the energy consumed by the attacker~\cite{Antoun2020}. 
Other traditional attacks that can be implemented on V2G networks are the information disclosure of the users' data, eavesdropping, impersonation attacks, and sybil attacks that were documented in~\cite{han2016privacy}.
The majority of these attacks are only theoretical and never implemented because of the difficulties in finding affordable testbeds.
However, since threats are real, it is therefore paramount to study and develop new security strategies to protect these infrastructures.



\parag{Contribution}In this paper, we present MiniV2G, a V2G cost-effective charging simulator tool to support the research on electric charging systems. MiniV2G is built on Mininet to support the network development while implementing RiseV2G to simulate the actual charging process based on Standard ISO 15118. MiniV2G works in a lightweight emulated scenario built on the top of Mininet. The virtualization approach allows to easily share the network configuration and the code to replicate the experiments.
Furthermore, Mininet flexibility allows the easy implementation of a charging scenario's topology, the analysis of the traffic exchange between all entities involved, and the implementation of attacks simulating malicious attempts.
In order to demonstrate MiniV2G security research effectiveness, we implemented and presented two attack scenarios in MiniV2G based on known vulnerabilities related to the V2G communication.

\parag{Organization}The present paper is organized as follows.
Section~\ref{sec:bakcground} briefly recalls the main useful concepts to convey the paper's main goal and introduces Mininet and RiseV2G, the two tool fundamental building blocks of MiniV2G. Section~\ref{sec:miniv2g} and Section~\ref{sec:experiments} present the implementation details and two experiments performed on MiniV2G, respectively. Section~\ref{sec:related} provides an overview of the related work, and finally, Section~\ref{sec:conclusion} covers the conclusions and proposes some insight into the future direction of this project.

\section{Background}\label{sec:bakcground}

In this section, we briefly recall the main concepts useful to understand the remainder of the paper. In particular, in Section~\ref{subsec:v2g_backgroung}, we introduce V2G communication, and in Section~\ref{subsec:iso15118}, we recall the ISO 15118, which defines the communication between the charging column and the vehicle.
We also introduce two open-source softwares upon which we build our emulator. In particular, Section~\ref{subsec:mininet} introduces Mininet, which is used to emulate the V2G architecture and manage the communication, and Section~\ref{subsec:risev2g} presents RiseV2G, which is employed to simulate the EV charging process.


\subsection{Vehicle-to-Grid (V2G)}\label{subsec:v2g_backgroung}

The charging process of EVs can be implemented in several methods. One of the main strategies is to plug them directly into the domestic electric grid when needed. However, due to the EV's high capacity batteries, the domestic power system cannot be efficient. Moreover, a user may require a power supply while not at home.


To overcome these limitations, in 2001, Kempton et al.~\cite{Kempton2001} introduced the Vehicle-To-Grid (V2G) charging system paradigm. V2G allows EVs to communicate with the smart grid, which actively manages the power delivered to the connected EVs. To authenticate the EV to the smart grid and receive the charging power, a user plugs the EV into Electric Vehicle Supply Equipments (EVSEs), which are called, less formally, charging columns.

Originally, V2G communication was uni-directional since the energy flows were directed uniquely from the power grid to the vehicle.
Recently, with the development of more sophisticated smart grids and chargers, the use of bi-directional V2G technology is growing~\cite{Yong2015}. The bi-directional transmission is made up of exchanging energy and information between the EV and the power grid and vice-versa. 
The exchange of information offers a wide variety of services, like the automatic request of energy during low load periods with a consequent reduction of energy cost. At the same time, the EVs can sell energy during peak hours if the smart grid requires it. Thanks to this strategy, a user can make an annual regulation profit of about 300-450\$ depending on the EV daily average usage time~\cite{Li2015}. 

\begin{figure}[bt]
\centering
\includegraphics[width=\columnwidth]{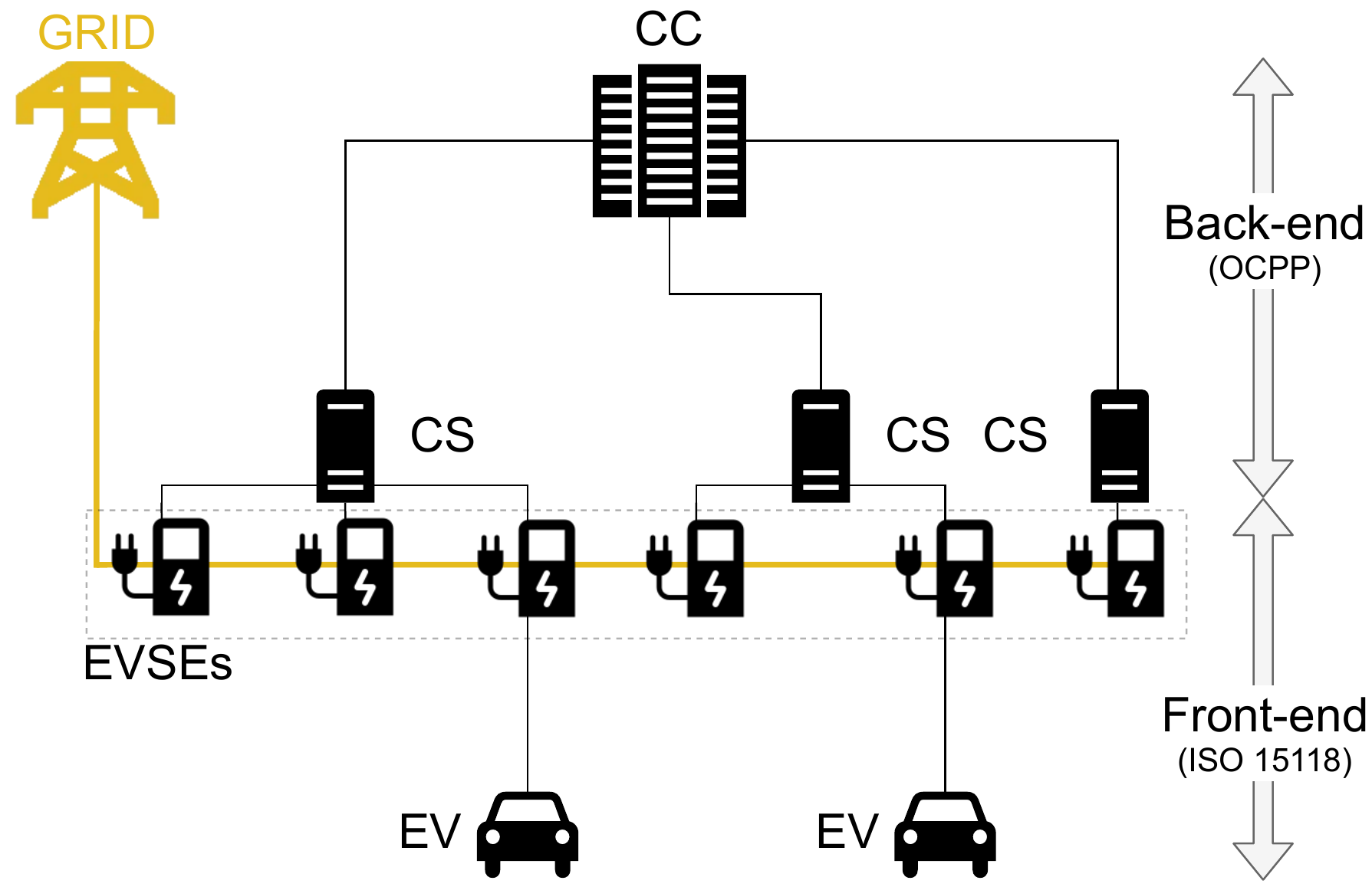}
\caption{The basic architecture of a V2G environment.}
\label{fig:v2g-architecture}
\end{figure}

The V2G infrastructure architecture, depicted in Figure~\ref{fig:v2g-architecture}, is composed of five entities: the EV, the Charging Stations (CSs), which is an aggregator of one or more EVSEs, the Control Center (CC) interconnecting points between CSs, and, finally, the smart grid to which all EVSEs are connected.

The communication in V2G environment~\cite{Schmutzler2013} is divided into two parts: front-end and back-end. The back-end communication regulates the interaction between the EVSE and the CC, generally occurring via the CS. More specifically, the back-end enables the EV to authenticate with the CC to recognize the user and charge the recharge fees. The back-end communication is defined in IEC 61850 standard. However, currently, the Open Charge Point Protocol (OCPP) developed by the Open Charge Alliance~\cite{ocpp2020} is the most widely used protocol and represents the de-facto standard. 
On the other hand, the front-end communication relies on the information shared between the EV and the EVSE. The two entities share information mainly on the EV State of Charge (SoC), the required charging parameters as well as the charging schedule. This also represents the primary means to reach the back-end. Although there are different protocols to regulate communication, the ISO 15118 standard implemented in MiniV2G is the most widely used.

\subsection{ISO 15118}\label{subsec:iso15118}

ISO 15118 regulates the communication in the V2G front-end between the EV communication component, i.e., Electric Vehicle Communication Controller (EVCC), and the EVSE communication component, i.e., Supply Equipment Communication Controller (SECC).
ISO 15118 is a recent standard: the development started in 2010, but it was officially published in 2013 and partially revised over the years, with the latest update in 2019. 

The main scope of this standard is to define the communication channel between the EV and the smart grid.
In future scenarios, billions of users will require energy to charge their EVs, so the grid load management will be fundamental to prevent overloads~\cite{6175685}. To this purpose, ISO 15118 allows a dynamic exchange of information between EV and EVSE to negotiate a charging schedule to match the user need and the grid load. For example, a user can connect the EV to a charger in the morning and require the charge to be completed in the evening.
In this case, the EV will not start the charge immediately by default. Instead, the smart grid will set the time-span for the EV to be charged, preferring the time slots with a low grid load and a lower energy price. Furthermore, if bi-directional V2G is enabled, the EV can sell energy to the grid during load peak hours, providing a profit for the EV owner.

\begin{figure}[bt]
\centering
\includegraphics[width=\columnwidth]{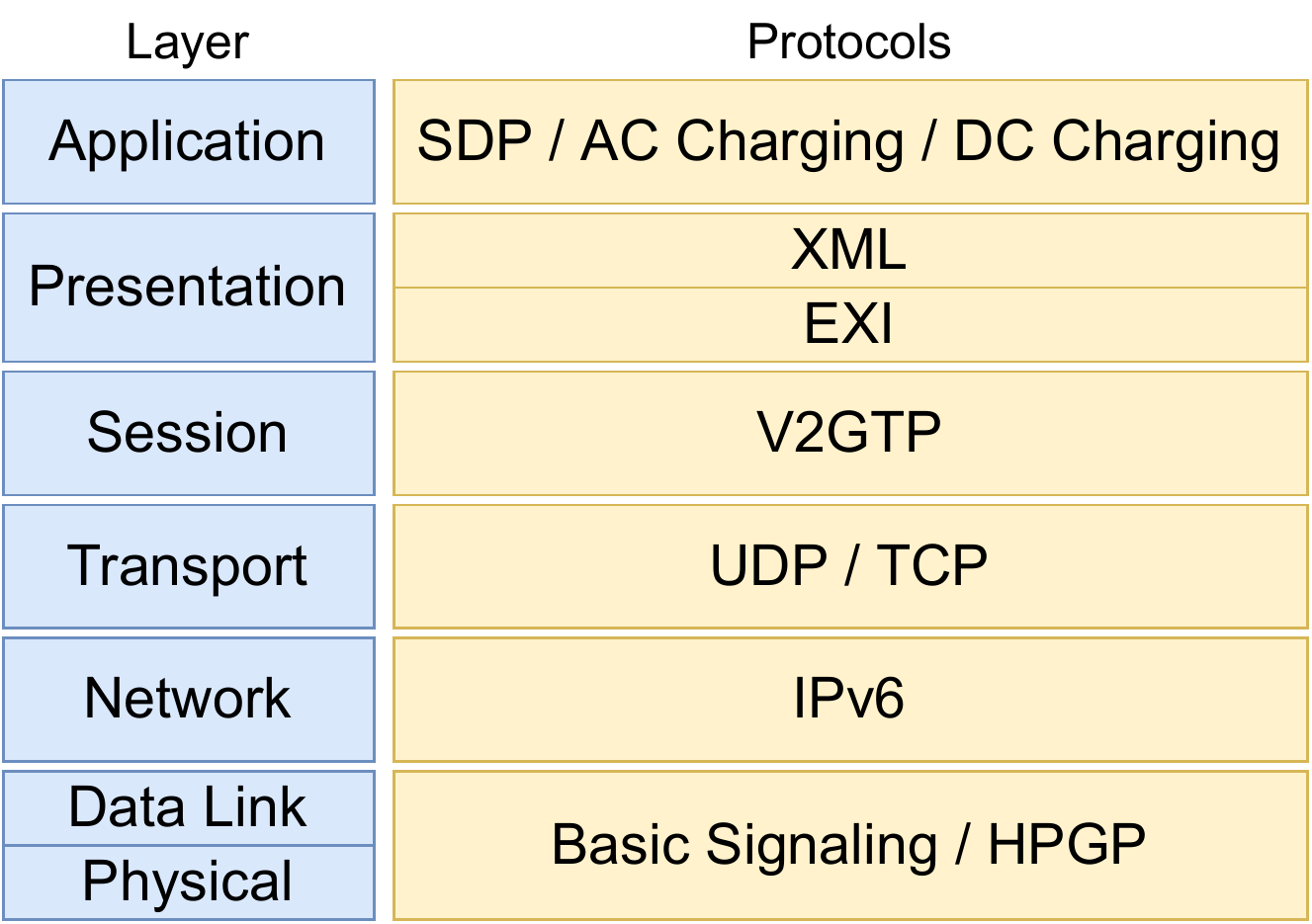}
\caption{Stack model of the ISO 15118 protocol compared to the seven ISO/OSI layers.}
\label{fig:stack}
\end{figure}

The layer stack structure of ISO 15118 is represented in Figure~\ref{fig:stack}, and each layer is presented in the following list:

\begin{itemize}
\item\textbf{Physical and Data Link Layer.} In the lower level, the Data Link and Physical layers are supported by two different technologies. A basic signaling protocol is employed, as standardized into IEC 61851, to perform the first synchronization phase. Afterward, Power Line Communication (PLC) can start following the HomePlug GreenPHY protocol (HPGP)~\cite{Alliance2010}. PLC is a technology used to transmit data on a cable, used simultaneously for electric power transmission. In this case, it uses the connector Control Pilot (CP) Pin.

\item\textbf{Network Layer.} The Network Layer implements the IPv6 suite since it offers a wider range of addresses with respect to IPv4. 

\item\textbf{Transport Layer.} At the Transport Layer, UDP, is used to exchange IPv6 addresses between EVCC and SECC employing the SECC Discovery Protocol (SDP) at the beginning of the communication. In SDP, the EVCC sends \texttt{SECCDiscoveryReq} as a broadcast message until the SECC replies sending its IPv6 address.
Then, a TCP session is established to send all the V2G messages to manage the charge. If sensitive data are involved, TLS can be adopted to guarantee a more secure exchange of information~\cite{Multin2018}.

\item\textbf{Session Layer.} The Session Layer handles communication sessions between the EV and the EVSE, employing a dedicated message Vehicle-to-Grid Transfer Protocol (V2GTP). The header of the message contains information on the payload type, stating whether it is an EXI encoded message type or an SDP message that is being transported~\cite{Multin2018}.

\item\textbf{Presentation Layer.} At the Presentation Layer, Extensible Markup Language (XML) packets from the Application Layer are compressed using the Efficient XML Interchange (EXI) for V2G messages. Compression is applied to all the V2G packets since they are in XML format, except for SDP messages.

\item\textbf{Application Layer.} Finally, the Application Layer header contains a Session ID. If the communication requires integrity and authenticity, which are mandatory requirements in a real network environment, an additional digital signature must be applied. This layer is responsible for V2G messages (both for AC and DC charging) and SDP messages. 

\end{itemize}

\begin{figure}[t]
\centering
\includegraphics[width=.8\columnwidth]{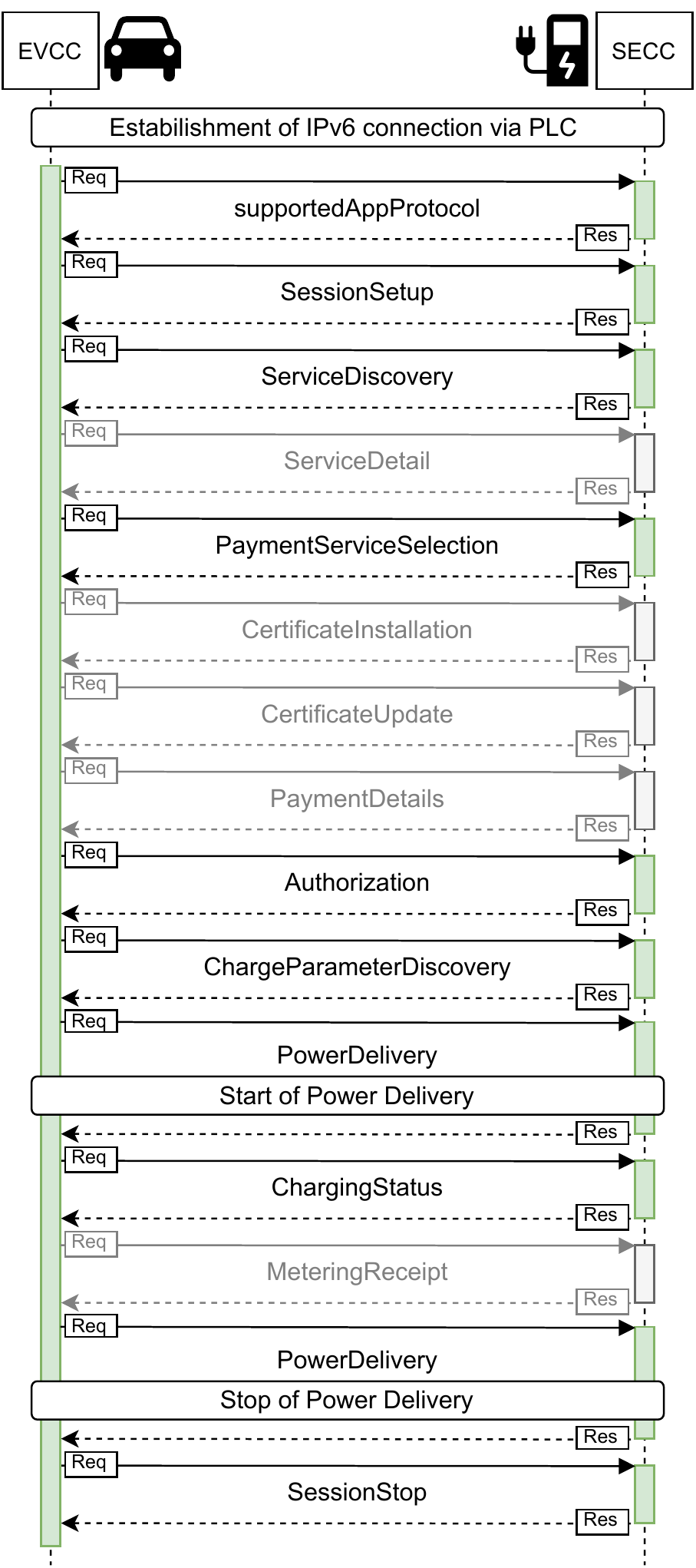}
\caption{The complete list of messages exchanged between the vehicle and a charging column during a charge. Optional messages are colored in gray.}
\label{fig:messages}
\end{figure}

Figure~\ref{fig:messages} summarizes the request/response message sequence to perform an EV charge~\cite{Multin2018}. Some messages are optional, for instance, \textbf{CertificateInstallation} and \textbf{CertificateUpdate} are exchanged only if TLS or Plug \& Charge is enabled. 

ISO 15118 introduces an important and user-friendly feature that allows the user to connect the EV and be automatically identified.  
This mode is called \emph{Plug \& Charge} and requires a back-end infrastructure to supply and manage certificates for the authentication phase. Another important service provided is Load Management, which helps the grid maintain load stability. 
Furthermore, the standard is open to the so-called Value Added Services that can be used by manufacturers or researchers to exploit the vehicle connected to the network and provide new services (e.g., the possibility to update the firmware of the vehicle's Electronic Control Units~\cite{Buschlinger2019}).

\subsection{Mininet}\label{subsec:mininet}

\emph{Mininet}~\cite{web:mininet} is an open-source software used to emulate realistic networks, running real kernel, switches, and application code. We chose Mininet to implement our system because it easily enables to build a virtual network topology and, if combined with other tools (e.g., Wireshark~\cite{web:wireshark}), it allows researchers and students to study and accurately reproduce customized network communications. Mininet also includes a graphical interface (i.e., Miniedit) to create network topology with a drag and drop manner. Furthermore, it is possible to interact with each virtual machine using the command line interface provided.
Mininet was originally built to enable and test Software-Defined Networking (SDN) scenarios, but it is suitable for many other use cases.

Over the years, \emph{Mininet} received numerous extensions to support different scenarios. One of the main extensions is \emph{Mininet-WiFi}~\cite{fontes2015mininet}, which enables emulating connection through WiFi connections. Our emulator is based on this version of Mininet to stimulate future development directions on the novel wireless power transfer technology applied to electric vehicles~\cite{sun2018review}.

\subsection{RiseV2G}\label{subsec:risev2g}

\emph{RiseV2G}~\cite{web:risev2g} is an open-source simulator for the ISO 15118 communication maintained by \emph{V2G Clarity}. RiseV2G includes all the ISO 15118-2 features for both AC and DC charging, including security-related features like encrypted communication via TLS and certificate handing to help developers set up Plug \& Charge. The software was tested at the biannual international Testing Symposium for ISO 15118 and the Combined Charging System (CCS) and resulted in a stable and reliable solution to create ISO 15118-compliant products. RiseV2G allows an easy functions setup thanks to the free online course provided by V2G Clarity~\cite{web:risev2g}.

RiseV2G is developed in Java and shared as a Maven project. It contains different code-including folders used to simulate EVCC and SECC. Moreover, the developers provide a folder with shared files and a helper to create certificates for the Public Key Infrastructure (PKI). 
RiseV2G, by default, provides only basic functionalities using the already developed simple controllers. 
However, RiseV2G is highly expansible since a complete simulation of a V2G environment can be implemented by extending the controllers.

\section{MiniV2G: our EVC emulator}\label{sec:miniv2g}


MiniV2G provides a testing environment for researchers working on V2G scenarios and for students to experiment with the V2G environment. 
In this section, we present MiniV2G emulator and its main functions. In particular, Section~\ref{subsec:arch} offers an architecture overview of our emulator, while Section~\ref{subsec:implementation} provides more in-depth implementation details of MiniV2G. 
Section~\ref{subsec:gui} presents the graphical interface, and lastly, Section~\ref{subsec:limitations} highlights the current limitations of our simulator.

\subsection{Architectural Overview}\label{subsec:arch}

\begin{figure}[bt]
\centering
\includegraphics[width=\columnwidth]{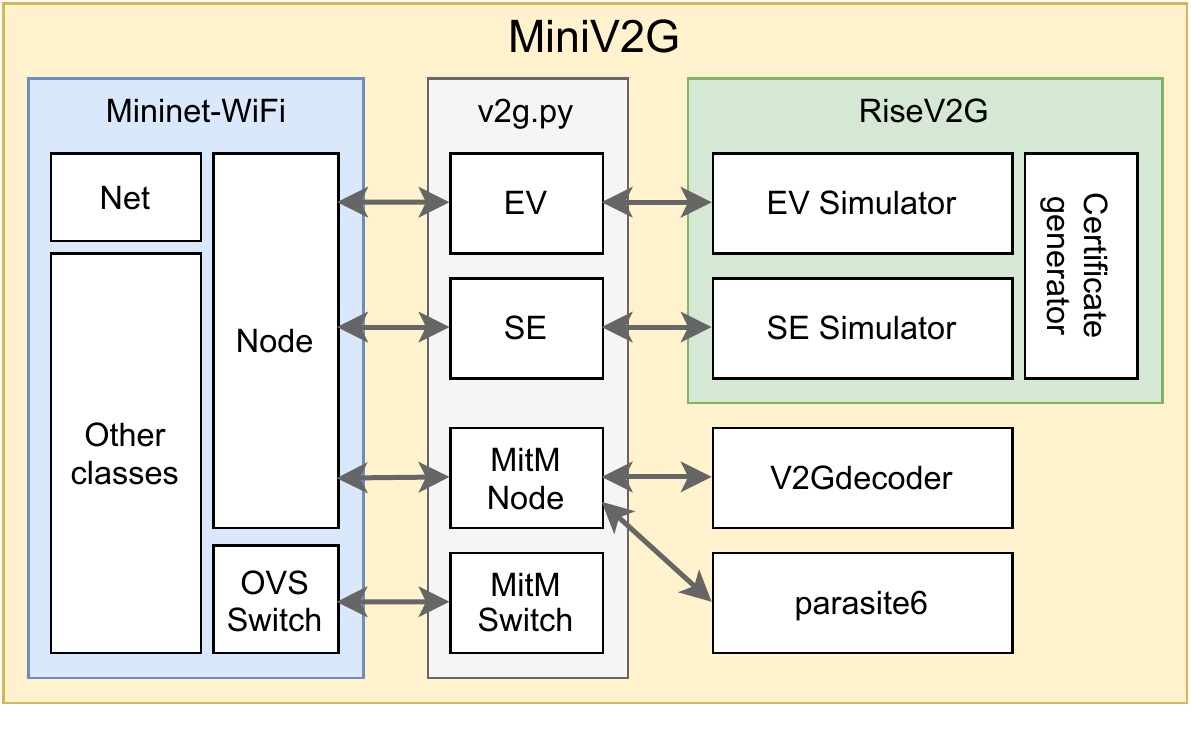}
\caption{The basic architecture of MiniV2G.}
\label{fig:arch}
\end{figure}

Mininet represents the core of the emulating part and offers SDN functions to create an inter-connected network of charging vehicles and charging columns. Along with the MiniV2G-specific entities, it is possible to employ the Mininet-WiFi suite of devices and components to provide wireless functionalities to the end-user. Instead, the charging process is simulated with RiseV2G. In MiniV2G, we maintain the basic functionalities provided by the RiseV2G \texttt{jar} files. 
Currently, the RiseV2G repository's includes the \texttt{jar} files of the 1.2.5 version. Since the most up to date RiseV2G version is instead the 1.2.6, we locally generate the \texttt{jar} files of such version that we uploaded as part of our repository. However, the installer of MiniV2G will check and download the latest version from the RiseV2G repository, and it will use the most recent one.
In addition, if any change to RiseV2G is necessary (e.g., implementation of a controller), it is possible to repack \texttt{jar} files and replace the default ones employed in MiniV2G.
The architecture of MiniV2G is summarized in Figure~\ref{fig:arch}, which shows the interconnections between Mininet-WiFi, the different tools, and the new classes.

To facilitate the implementation of Man-in-the-Middle (MitM) attacks, we employed two additional tools named parasite6~\cite{web:thc} and V2Gdecoder~\cite{web:v2gdecoder, Dudek2019}. We embedded the functionalities of these tools on a new node: the MitM Node. Furthermore, we leveraged the functions provided by OVSSwitch in Mininet by introducing the MitM Switch node, which specifically redirects the communication flows to the MitM Node.

\subsection{Implementation}\label{subsec:implementation}

MiniV2G is build on top of Mininet-WiFi~\cite{fontes2015mininet} emulator and integrates RiseV2G~\cite{web:risev2g} to simulate the charging process. The source code of the simulator is available on the GitHub repository~\footnote{\url{https://www.github.com/donadelden/miniV2G/}. Note: at the submission time, the commit code is d827168}, and it is completely open-source. The simulator architecture is based on Mininet-WiFi with two new entities to implement the V2G communications. 
The two new classes enabling V2G communications through the RiseV2G simulator are the following:
\begin{itemize}
    \item\textbf{SE (Supply Equipment)} is a host expansion class providing EVSE functionalities such as starting to listen for EVs which require a charge and to change or get the initialized station parameters;
    \item\textbf{EV (Electric Vehicle)} is a class designed to expand a host by providing a function to charge the EV and to set/get the vehicle parameters.
\end{itemize}
Once the architecture is designed, with the initialization functions of MiniV2G, the RiseV2G \texttt{jar} simulator files and its properties files are copied into an ad-hoc folder accessible by the host. The corresponding host configuration file can be modified manually or by using the API provided in the code. 
MiniV2G allows to configure the following parameters on the EV.
\begin{itemize}
    \item\texttt{voltage.accuracy}: to set the accuracy in the measurement of energy;
    \item\texttt{tls}: to activate the TLS communication;
    \item\texttt{session.id}: to set the charge Session ID;
    \item\texttt{network.interface}: indicating the interface to communicate with other devices;
    \item\texttt{energy.transfermode.requested}: indicating the charging modes required (e.g., AC single/three phase, DC extended)
\end{itemize}
Furthermore, the SE has some important properties, which can be set as follows.
\begin{itemize}
    \item\texttt{free.service}: indicating if the charging column requires a payment;
    \item\texttt{network.interface}: indicating the interface to communicate with other devices;
    \item\texttt{energy.transfermodes.supported}: indicating the supported charging modes (e.g., AC single/three phase, DC extended).
\end{itemize}
Other properties can be modified to change some parameters, and a complete list is provided on the MiniV2G repository documentation.

To facilitate the security research with MiniV2G and implementing attack scenarios, we developed a MitM node. The node targets the communication channel between the EV and SE. Furthermore, we also implemented a packet decoder to manipulate the EXI version of the communication.

The MitM node is implemented with two additional classes:
\begin{itemize}
    \item\textbf{MitM Switch} is a class designed to expand an Open vSwitch (OVS) and add flows to the switch, which redirects the traffic to a MitM Node. More precisely, this node uses the SDN function, which was already integrated by Mininet, to establish a direct communication link for the attacker to intercept and send messages to the EV or SE.
    \item\textbf{MitM Node} is a host expansion class including functions to perform IPv6 neighbor spoofing via parasite6~\cite{web:thc}, to start a web-server that forwards messages from the SE to the EV or a web-server which will perform a DoS on the communication channel between an SE and an EV. This class also includes a method to start a web-server version of V2Gdecoder~\cite{web:v2gdecoder, Dudek2019}, a decoder for EXI messages;
\end{itemize}

\subsection{Graphical Interface}\label{subsec:gui}

An interesting tool for Mininet and Mininet-WiFi is Miniedit, a practical graphical interface that can be employed to design, configure, and test basic topologies in a drag-and-drop manner to support the user on the first steps when using the emulator. MiniV2G offers an expanded user interface, enabling to insert an EV and EVSE to the network. Furthermore, MiniV2G includes a dedicated window to configure the RiseV2G nodes properties, according to the properties introduced in Section~\ref{subsec:implementation}.
MiniV2G also allows the creation of the desired network topology via code. If the reader prefers this approach, command-line examples are also provided in the \texttt{examples} folder of MiniV2G.

\begin{figure}[bt]
\centering
\includegraphics[width=0.8\columnwidth]{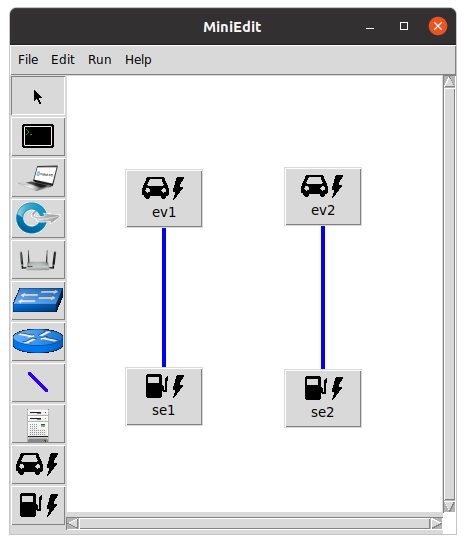}
\caption{An example of graphical interface tool-generated simple topology.}
\label{fig:gui}
\end{figure}

Figure~\ref{fig:gui} shows the graphical interface of MiniV2G with a basic topology containing two EVs with two interconnected SEs. MiniV2G allows implementing more complex scenarios, with more than one EVs connected to a single SE. Furthermore, it is possible to connect other types of nodes to an EV or an SE (e.g., implementing a back-end infrastructure).
The user interface is the simplest way to create a project and is advisable for entry-level users. From the interface, loading and exporting topologies is possible. This feature is useful to share pre-configured networks with other researchers, for example, to enable the reproduction of an experiment.
It is also possible to generate a standalone Python script that can be manually modified to add functionalities and fulfill more sophisticated tasks. This latter case is more suitable if the user needs to gain a better insight into the protocol.
The exported code contains the simplest way to create a network with EVs and SEs and represents the skeleton to build the required project.
To include either an SE or EV V2G entity, the \texttt{addNode()} function of the \texttt{net} class must be used and the parameter \texttt{cls} has to be set to \texttt{EV} or \texttt{SE} respectively.

\subsection{Current Limitations}\label{subsec:limitations}

Currently, MiniV2G allows simulating the charging process only at network-level communication.
In fact, a complete study of the charging process may require information on physical-level parameters, such as the true voltage which an EVSE is delivering. RiseV2G does not provide this functionality, and, to the best of our knowledge, no software offers a complete simulation of ISO 15118. Furthermore, the charging process lasts for one charging cycle, which in RiseV2G results in few seconds of operation. Moreover, currently, RiseV2G does not allow to set the charging duration.

\section{Example Application}\label{sec:experiments}

In this section, we present two experiments performed on MiniV2G to show its functioning. In particular, in Section~\ref{subsec:system_model} we focus on the MitM system scenario exploited to perform the attacks. In Section~\ref{subsec:attacks} we show the implementation of a Message Modification and a DoS attack, while in Section~\ref{subsec:counter} we discuss the protection mechanism to prevent the presented attacks.



\subsection{Designed System Scenario}\label{subsec:system_model}

\begin{figure}[bt]
\centering
\includegraphics[width=\columnwidth]{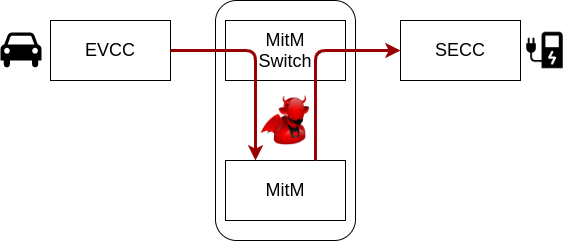}
\caption{MitM architecture in MiniV2G. The MitM switch redirects the flow to MitM, which intercepts and modifies or forwards the packet to SE.}
\label{fig:mim_architecture}
\end{figure}

We designed the system scenario to conduct the study on the attacks on a classic charging process between EV and SE. In particular, a user who needs to recharge its EV, connect it to the SE, and start the charging process. 
In the system model, however, we assume that an attacker could compromise the communication channel, for example, by placing an additional device between the EV and SE, interfering on the connection cable. This enables the attacker to sniff and modify the communication. This operation might be done at night when there are no people around the CS.
In Figure~\ref{fig:mim_architecture}, the different network entities of the highlighted scenario are depicted. The device plugged in by the attacker is a combination of the MitM Switch and MitM Node, as explained in Section~\ref{subsec:implementation}.
The MitM Switch redirects the network flow from the MitM Switch to the MitM Node, which then decides if it has to forward, interrupt, modify or fabricate a new packet to the destination.
Despite V2G allows the implementation of TLS to secure the communication, different studies highlight that, currently, on practical scenarios, this security measure is often not implemented~\cite{baker2019losing}, falling back to a TCP communication and exposing the channel to classical eavesdropping attacks. Furthermore, the same article~\cite{baker2019losing} explains that when an external mean of payment is employed, which is the case of the majority of the charging columns up to date, TLS is not mandatory. For these reasons in our system scenario, we assume that the communication between EVCC and SECC is not encrypted by default, resulting in a system that we proved to be insecure in Section~\ref{subsec:attacks}.

\subsection{Simulated Attacks}\label{subsec:attacks}

To facilitate the implementation of such attacks, in MiniV2G, we implemented a MitM node that can be placed between an EV and the SE (Figure~\ref{fig:mim_architecture}) to interrupt, modify, fabricate or intercept messages exchanged between the two parties.
In MiniV2G, the MitM Switch was implemented with the aid of Open vSwitch~\cite{pfaff2015design}, a multi-layer virtual switch, configured to redirect network flows to the MitM Node, and parasite6, part of THC-IPv6 tools~\cite{web:thc}, a spoofer executed by the MitM to redirect all traffic by answering with the MitM MAC address to Neighbour Solicitation requests. 
The MitM node runs a web-server that redirects or modifies the UDP or TCP messages sent from the EV to the SE and viceversa. In addition, the MitM Node implements V2Gdecoder~\cite{web:v2gdecoder, Dudek2019} to facilitate the decoding of the EXI payload exchanged between the EV and the SE or to encode the XML representation of a V2G message. 

Once the attacker is able to redirect the entire communication to the MitM Node, full control of the communication is gained and different types of attacks can be performed.

\parag{Message Modification} As previously explained in Section~\ref{subsec:iso15118}, the SDP employs UDP messages to exchange the IPv6 addresses and ports of the EV and SE. In this attack, the EV broadcasts a UDP message and the MitM Node modifies the UDP message replies from the SE before forwarding it back to the EV, changing the SE port to a new port linked to its malicious web-server. The MAC address faking is implemented with parasite6. This tool enforces the attack, which could also be conducted by changing the IPv6 of the SE in the UDP message.
Once SDP is over, V2G messages will be exchanged using TCP from the start to the end of the V2G session in a request-reply manner from the EV to the SE. In this communication, the order is important and the EV always acts as an initiator, waiting for a reply from the SE.

\begin{figure}[bt]
\centering
\includegraphics[width=\columnwidth]{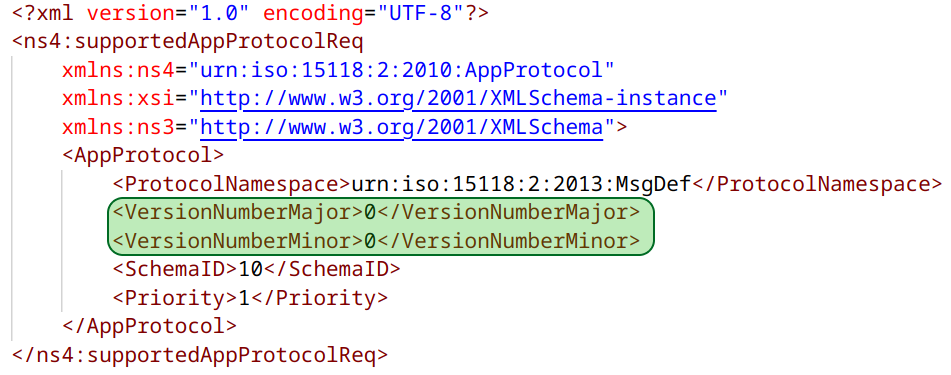}
\caption{The DoS attack modifies the content of \textbf{supportedAppProtocolRequest}, by changing the EV version of the protocol to 0.0, which is not supported by the SECC.}
\label{fig:supported_app_xml}
\end{figure}

\parag{Denial of Service} 
MiniV2G currently supports a DoS attack that exploits the MitM node to modify the content of the \textit{supportedAppProtocolRequest} message. This message is a request to agree on the protocol version sent from the EV to the SE. The goal of the DoS attack is to interrupt the communication between the EVCC and the SECC by blocking the charging process. 
To do this, the attack sets the protocol version supported by the EV to a version unsupported by RiseV2G (e.g., \textit{0.0}), making the SE service unavailable to the EV. We targeted this particular message because it is the most vulnerable of the series to a DoS attack since the Session ID has not been set yet. In fact, every other V2G message contains a Session ID in the application layer header and it is required to be known before altering them. 
We reported in Figure~\ref{fig:supported_app_xml} an XML modification to perform the DoS attack.


The MitM node can be implemented to perform other malicious operations by altering the EXI payload of the messages exchanged between the EV and SE, to study the consequences of such modification, and to develop protection mechanisms. After the session setup, all the V2G messages are exchanged with a header containing a Session ID, and the attacker may spoof the Session ID from the \textit{SessionSetupRequest} to alter them.
By exploiting the MitM node, an attacker can also replace the \textit{ServiceDiscoveryResponse} to add or remove charging services supported by the SE or the \textit{ServiceDiscoveryResponse} to deny support for an authentication and authorization method.
Furthermore, by modifying the \textit{PowerDeliveryResponse} and \textit{SessionStopRequest} messages, the attacker can pause, stop or reschedule the EV charge, beforehand.

\subsection{Countermeasure}\label{subsec:counter}

To study the application of countermeasures to attacks such as MitM and DoS, MiniV2G implements TLS to provide confidentiality and data integrity in the communication between EVs and SE. 
If TLS is enabled and an attacker tries to perform a MitM between the EV and SE, the MitM will cause the TLS handshake to fail. As a consequence, the user might be aware of a connection problem. To activate TLS, it is possible to use the \texttt{setTLS} function of the \texttt{EV} class. This function integrates the generation of TLS certificates in RiseV2G to set up a TLS connection between an EV and a SE.
Thanks to the real-world charging process fidelity of MiniV2G, the study and the implementation of encryption techniques such as TLS in the simulated scenario can be easily transposed in a real-world application.
Furthermore, the possibility of implementing encrypted communication such as TLS allows the researcher to perform studies on the encrypted traffic analysis.







\section{Related Works}\label{sec:related}
In this section, we present different works that extend Mininet to satisfy different researchers' needs. Then, we recall other analyses of the V2G environment's security, highlighting the few implemented attacks in literature. Furthermore, we present an overview of the V2G simulators and testbeds available.

\parag{Mininet Extensions} Mininet is a popular network emulator used to create several ad-hoc emulations to meet different researchers' needs. An example is the already mentioned \emph{Mininet-WiFi}~\cite{fontes2015mininet}, which enables wireless communication between entities. \emph{Mininet-WiFi} also provides an entity to simulate cars and Vehicular ad hoc networks (VANETs), but without taking into account electric vehicles and their charging communication. MiniCPS~\cite{Antonioli2015mini} is another popular extension of Mininet developed as a toolkit specifically designed to support security research in the field of Cyber-Physical Systems (CPSs), Industrial Control Systems, and critical infrastructures. 
MiniCPS provides, along with Mininet real-time network emulation ability,  emulation scripts for CPS components (e.g., Programmable logic controllers, sensors, actuators) and a simple API to interface with real-time physical-layer emulation. 
Other extensions supporting different features have been presented, for instance, \emph{Minievents}~\cite{web:minievents} adds an event generator to create network changes at specific times, while \emph{Containernet}~\cite{web:containernet} makes Mininet able to create hosts from Docker containers.


\parag{V2G Attacks} Several works in the literature discuss the possibility of implementing attacks on the V2G environment~\cite{han2016privacy, Mustafa2013}. Antoun et al.~\cite{Antoun2020} presented a theoretical survey highlighting possible attacks targeting different players in the charging scenario and the countermeasures proposed in the literature. Other studies focus instead on a particular target, for instance, on the channel between the charging piles and the electric vehicles~\cite{Xu2019} or the security of the charging station~\cite{Gottumukkala2019}. However, the majority of the studies are limited on the attack presentation, without a real implementation. There are only a few works that effectively implement attacks on a real scenario. Dudek et al.~\cite{Dudek2019} proposed V2GInjector~\cite{web:v2ginjector}, a tool they employ to exploit a specification vulnerability in the communication medium on the V2G network.
Instead, in~\cite{baker2019losing} the authors were able to sniff the Media Access Control (MAC) address from unshielded charging cables employed in the connections between EVSE and EV. They implemented the attack on several different charging stations in England to show the impact of the research. 

The lack of attack implementations in the V2G field is mainly motivated by the sector's expensiveness and closeness, where standards are generally available for sale only and the required equipment (e.g., EV, EVSE) is costly and not accessible for everyone. To overcome these limitations, our emulator supports researchers to study, test attacks and countermeasures in this field. Moreover, many of the theoretical authentication schemes proposed in literature~\cite{saxena2016authentication, garg2019efficient, su2019novel, bansal2020lightweight}, proved to be privacy-preserving by the authors, could be verified with our emulator.

\parag{Physical V2G Testbeds} Some efforts have been made towards the development of physical testbeds to simulate the V2G environment. 
Dudek et al.~\cite{Dudek2019} show the possibility to build a cheap system to interface with V2G using a Devolo development kit~\cite{web:devkit} based on a QCA7000 (QCA7k) module by Qualcomm Atheros. The presented testbed provides interfaces for a twisted pair coax cable and an AC line. On the other side, it is equipped with an Ethernet port that can be used to set up the PLC modem. The authors estimate the cost of the testbed between 150-200\euro{}. Moreover, the authors presented some useful tools for the community: V2Ginjector~\cite{web:v2ginjector}, a software to penetrate V2G networks, monitor and inject packets to attack electric cars and to charge stations, and V2Gdecoder~\cite{web:v2gdecoder}, a tool to decode EXI packets which is also used into MiniV2G. However, physical testbeds may not be easily accessible to anybody. Furthermore, the replication and verification of the experiment could raise concerns.

\parag{V2G Simulators} In~\cite{Faschang2015}, Faschang et al. proposed a co-simulation framework to test interoperability on heterogeneous compounds of charging infrastructures. They present different approaches to build this framework to meet several researchers' needs. For instance, self-defined protocols can be adopted and interconnected with real-word hardware using protocol converters.
However, this work mainly focuses on the back-end part of the communication employing protocols such as OCPP, without considering the front-end communication.
In the field of EVs, there are no physical emulators of ISO 15118 communication. 
To the best of the authors' knowledge, only two simulators are available: RiseV2G and OpenV2G. We chose to include RiseV2G in our project because it is open source, and it provides a free course to understand the basic functionalities of the software.
OpenV2G~\cite{web:openv2g} is another relevant project implemented in C and developed by Siemens. OpenV2G was born as an open-source project, but, as specified in~\cite{web:openv2g}, from ISO 15118-2-2016 standard (ISO Edition 2 of 2016), the project is no longer open-source. The last public version is dated 2018, and no further updates will be provided. However, the modularity of MiniV2G allows supporting OpenV2G as a replacement of RiseV2G, if needed.

\section{Conclusion}\label{sec:conclusion}

In this work, we proposed MiniV2G, a tool to emulate a V2G architecture, including the EV charging process, to support the V2G field research. MiniV2G's main goal is to provide researchers a cost-effective testbed with a high degree of fidelity. Furthermore, MiniV2G's virtualized nature enables the topologies exportation to replicate the experiments and validate them. 

To demonstrate MiniV2G functioning, we implemented a security-oriented application study based on two attacks, which exploit ISO 15118 vulnerabilities, that malicious users could implement in a real-world charging scenario.
We openly disclosed the source code of MiniV2G to share the project with the community and make it open-source for future development and suggestions.

\parag{Future Works} In the future works, we plan to extend the implementation of TLS by building a full PKI to support the management of certificates. This feature can enable automatic download and renewal of EV certificates and simulate contracts for Plug \& Charge associated with load balance.
Furthermore, we aim to extend MiniV2G to simulate Charging-When-Driving (CWD) scenarios by resorting to the host movement functionalities of Mininet-WiFi.



\begin{acks}
We thank the authors of Mininet, Mininet-WiFi, and RiseV2G for sharing the source code of their projects.
\end{acks}

\bibliographystyle{ACM-Reference-Format}
\bibliography{bibliography}



\end{document}